\documentclass{Rinton-P10x7}

\input{psfig}
\usepackage{epsfig}
\usepackage{pstricks}

\def\slash#1{\rlap{\hbox{$\mskip 1 mu /$}}#1} 
\def\pa{\partial} 
\def\slpa{\slash{\pa}}                            

\def\a{\alpha}
\def\b{\beta}

\def\d{\delta}
\def\e{\epsilon}
\def\f{\phi}
\def\g{\gamma}
\def\h{\eta}

\def\j{\psi}
\def\k{\kappa}
\def\l{\lambda}
\def\m{\mu}
\def\n{\nu}

\def\p{\pi}

\def\r{\rho}
\def\s{\sigma}
\def\t{\tau}

\def\z{\zeta}
\def\D{\Delta}
\def\F{\Phi}
\def\G{\Gamma}

\def\L{\Lambda}

\def\P{\Pi}
\def\Q{\Theta}
\def\S{\Sigma}

\def\ve{\varepsilon}
\def\vf{\varphi}

\def\co{{\cal O}}

\def\cl{{\cal L}}

\def\pa{\partial}                                       
\def\der#1{{\pa \over \pa {#1}}}                
\def\dg{\dagger}                                     
\def\VEV#1{\left\langle #1\right\rangle}        
\def\Bar#1{\overline{#1}}                       
\def\beq{\begin{equation}}
\def\eeq{\end{equation}}
\def\bea{\begin{eqnarray}}
\def\eea{\end{eqnarray}}
\def\NO{\nonumber}

\def\pl#1#2#3{Phys.~Lett.~{\bf B {#1}} ({#2}) #3}
\def\np#1#2#3{Nucl.~Phys.~{\bf B {#1}} ({#2}) #3}
\def\prl#1#2#3{Phys.~Rev.~Lett.~{\bf #1} ({#2}) #3}
\def\pr#1#2#3{Phys.~Rev.~{\bf D {#1}} ({#2}) #3}

\def\prep#1#2#3{Phys.~Rep.~{\bf {#1}C} ({#2}) #3}

\begin{document}

\title{
\vspace{-.5cm}
{\rm\normalsize DESY 01-101 \hfill\mbox{ }}\\[5ex]
Recent Progress in Leptogenesis\footnote{Talk presented at PASCOS 2001, 
Chapel Hill, USA}}

\author{W. Buchm\"uller}

\address{Deutsches Elektronen-Synchrotron DESY, Hamburg, Germany}


\maketitle

\abstracts{
After recalling the general virtues of leptogenesis we compare two
realizations, Affleck-Dine leptogenesis and thermal leptogenesis,
which generically lead to different predictions for neutrino masses.
Finally, we describe the progress towards a full quantum mechanical 
description of the basic non-equilibrium process of leptogenesis
beyond the approximations involved in Boltzmann's equations.} 

\section{Why Leptogenesis ?}

One of the main successes of the standard early-universe cosmology is the
prediction of the abundances of the light elements, D, $^3$He, $^4$He and 
$^7$Li. Agreement between theory and observation is obtained for
a certain range of the parameter $\eta$, the ratio of baryon density and
photon density\cite{rpp00},
\beq
\eta = {n_B\over n_\g} = (1.5 - 6.3)\times 10^{-10}\;,
\eeq
where the present number density of photons is $n_\g \sim 400/{\rm cm}^3$. 
Since no significant amount of antimatter is observed in the universe, 
the baryon density yields directly the cosmological baryon asymmetry, 
$Y_B =(n_B - n_{\bar{B}})/s \simeq \eta/7$, where $s$ is the entropy density.

A matter-antimatter asymmetry can be dynamically generated in an expanding
universe if the particle interactions and the cosmological evolution satisfy 
Sakharov's conditions\cite{sa67},i.e., 
\begin{itemize}
\item baryon number violation
\item $C$ and $C\!P$ violation
\item deviation from thermal equilibrium .
\end{itemize}
Although the baryon asymmetry is just a single number, it provides an
important relationship between the standard model of cosmology, i.e., the
expanding universe with Robertson-Walker metric, and the standard model
of particle physics as well as its extensions.

At present there exist a number of viable scenarios for baryogenesis. They
can be classified according to the different ways in which Sakharov's 
conditions are realized. Already in the standard model $C$ and $C\!P$ are
not conserved. Also baryon number ($B$) and lepton number ($L$) are
violated by instanton processes\cite{tho76}. In grand unified theories
$B$ and $L$ are broken by the interactions of gauge bosons and leptoquarks.
This is the basis of the classical GUT baryogenesis\cite{yo78}.
Analogously, the $L$-violating decays of heavy Majorana neutrinos lead to
leptogenesis\cite{fy86}. The initial abundance of the heavy neutrinos may
be generated either thermally or from inflaton decays\cite{ahx00,jkx00}.
In supersymmetric theories the existence of 
approximately flat directions in the scalar potential allows for new 
possibilities. Coherent oscillations of scalar fields may then generate
large asymmetries\cite{ad85}.

The crucial departure from thermal equilibrium can also be realized in several
ways. One possibility is a sufficiently strong first-order electroweak phase 
transition\cite{rt99}. In this case $C\!P$ violating interactions of the
standard model or its supersymmetric extension could in principle generate the
observed baryon asymmetry. However, due to the rather large lower bound on the
Higgs boson mass of about 115~GeV, which is imposed by the LEP experiments,
this interesting possibility is now restricted to a very small range of 
parameters in the supersymmetric standard model. In the case of the 
Affleck-Dine scenario the baryon asymmetry is generated at the end of an 
inflationary period as a coherent effect of scalar fields which leads to an
asymmetry between quarks and antiquarks after reheating\cite{drt96}.
For the classical GUT baryogenesis and for leptogenesis the departure from
thermal equilibrium is due to the deviation of the number density of the
decaying heavy particles from the equilibrium number density.
How strong this deviation from thermal equilibrium is depends on the lifetime
of the decaying heavy particles and the cosmological evolution. Further
scenarios for baryogenesis are described in ref.\cite{dol92}.

The theory of baryogenesis involves non-perturbative aspects of quantum
field theory and also non-equilibrium statistical field theory, in particular
the theory of phase transitions and kinetic theory. 
\begin{figure}[h]
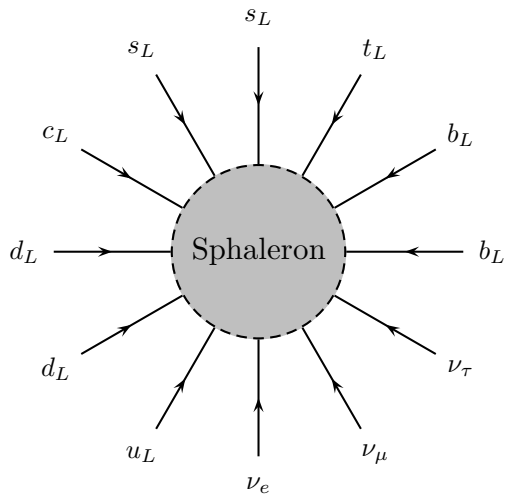

\begin{center}
\scaleboxto(7,7) {\parbox[c]{9cm}{ \begin{center}
     \pspicture*(-0.50,-2.5)(8.5,6.5)
     \psset{linecolor=lightgray}
     \qdisk(4,2){1.5cm}
     \psset{linecolor=black}
     \pscircle[linewidth=1pt,linestyle=dashed](4,2){1.5cm}
     \rput[cc]{0}(4,2){\scalebox{1.5}{Sphaleron}}
     \psline[linewidth=1pt](5.50,2.00)(7.50,2.00)
     \psline[linewidth=1pt](5.30,2.75)(7.03,3.75)
     \psline[linewidth=1pt](4.75,3.30)(5.75,5.03)
     \psline[linewidth=1pt](4.00,3.50)(4.00,5.50)
     \psline[linewidth=1pt](3.25,3.30)(2.25,5.03)
     \psline[linewidth=1pt](2.70,2.75)(0.97,3.75)
     \psline[linewidth=1pt](2.50,2.00)(0.50,2.00)
     \psline[linewidth=1pt](2.70,1.25)(0.97,0.25)
     \psline[linewidth=1pt](3.25,0.70)(2.25,-1.03)
     \psline[linewidth=1pt](4.00,0.50)(4.00,-1.50)
     \psline[linewidth=1pt](4.75,0.70)(5.75,-1.03)
     \psline[linewidth=1pt](5.30,1.25)(7.03,0.25)
     \psline[linewidth=1pt]{<-}(6.50,2.00)(6.60,2.00)
     \psline[linewidth=1pt]{<-}(6.17,3.25)(6.25,3.30)
     \psline[linewidth=1pt]{<-}(5.25,4.17)(5.30,4.25)
     \psline[linewidth=1pt]{<-}(4.00,4.50)(4.00,4.60)
     \psline[linewidth=1pt]{<-}(2.75,4.17)(2.70,4.25)
     \psline[linewidth=1pt]{<-}(1.83,3.25)(1.75,3.30)
     \psline[linewidth=1pt]{<-}(1.50,2.00)(1.40,2.00)
     \psline[linewidth=1pt]{<-}(1.83,0.75)(1.75,0.70)
     \psline[linewidth=1pt]{<-}(2.75,-0.17)(2.70,-0.25)
     \psline[linewidth=1pt]{<-}(4.00,-0.50)(4.00,-0.60)
     \psline[linewidth=1pt]{<-}(5.25,-0.17)(5.30,-0.25)
     \psline[linewidth=1pt]{<-}(6.17,0.75)(6.25,0.70)
     \rput[cc]{0}(8.00,2.00){\scalebox{1.3}{$b_L$}}
     \rput[cc]{0}(7.46,4.00){\scalebox{1.3}{$b_L$}}
     \rput[cc]{0}(6.00,5.46){\scalebox{1.3}{$t_L$}}
     \rput[cc]{0}(4.00,6.00){\scalebox{1.3}{$s_L$}}
     \rput[cc]{0}(2.00,5.46){\scalebox{1.3}{$s_L$}}
     \rput[cc]{0}(0.54,4.00){\scalebox{1.3}{$c_L$}}
     \rput[cc]{0}(0.00,2.00){\scalebox{1.3}{$d_L$}}
     \rput[cc]{0}(0.54,0.00){\scalebox{1.3}{$d_L$}}
     \rput[cc]{0}(2.00,-1.46){\scalebox{1.3}{$u_L$}}
     \rput[cc]{0}(4.00,-2.00){\scalebox{1.3}{$\nu_e$}}
     \rput[cc]{0}(6.00,-1.46){\scalebox{1.3}{$\nu_{\mu}$}}
     \rput[cc]{0}(7.46,0.00){\scalebox{1.3}{$\nu_{\tau}$}}
     \endpspicture
\end{center}}}
\end{center}
\caption{One of the 12-fermion processes which are in thermal 
equilibrium in the high-temperature phase of the standard model.
\label{fig_sphal}}
\end{figure}
A crucial ingredient is the connection between baryon number and lepton 
number in the high-temperature, symmetric phase of
the standard model. Due to the chiral nature of the weak interactions $B$ and
$L$ are not conserved. At zero temperature this has no observable 
effect due to the smallness of the weak coupling. However, as the temperature 
approaches the critical temperature $T_{EW}$ of the electroweak transition, 
$B$ and $L$ violating processes come into thermal equilibrium\cite{krs85}. 

The rate of these processes is
related to the free energy of sphaleron-type field configurations which carry
topological charge. In the standard model they lead to an effective
\begin{figure}[t]
\begin{center}
\epsfig{file=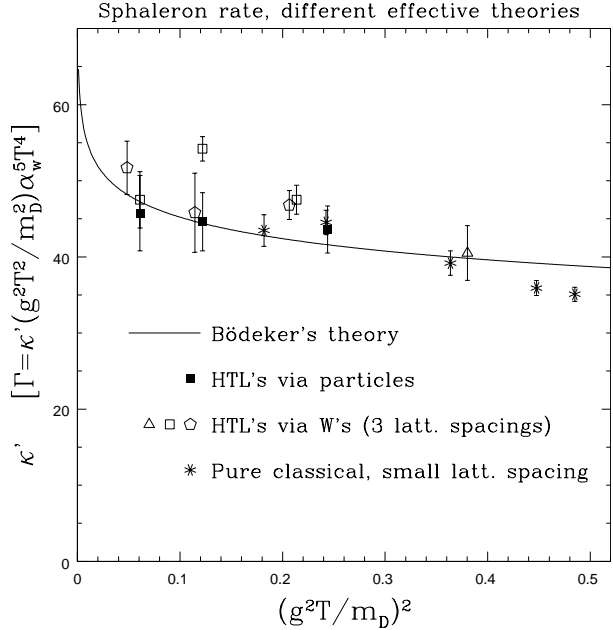,width=8cm,clip=}
\caption{Sphaleron rate in B{\"o}deker's effective theory, two lattice 
implementations of the HTL effective theory, and pure lattice theory interpreted as HTL effective theory\protect\cite{moo00}.}
\label{boed_fig}
\end{center}
\end{figure}
interaction of all left-handed quarks and leptons\cite{tho76} 
(cf. fig.~\ref{fig_sphal}), 
\begin{equation}\label{obl}
O_{B+L} = \prod_i \left(q_{Li} q_{Li} q_{Li} l_{Li}\right)\; ,
\end{equation}
which violates baryon and lepton number by three units, 
\begin{equation} 
    \D B = \D L = 3\;. \label{sphal1}
\end{equation}

The evaluation of the sphaleron rate in the symmetric high-temperature phase 
is a complicated problem. A clear physical picture has been obtained in
B\"odeker's effective theory\cite{boe98} according to which low-frequency 
gauge field fluctuations satisfy the equation of motion
\beq
{\bf D}\times {\bf B} = \s {\bf E} - {\bf \z}\; .
\eeq
Here $\bf \z$ is Gaussian noise, i.e., a random vector field with variance
\beq
\langle \z_i(t,{\bf x})\z_j(t',{\bf x'})\rangle = 
2\s \d_{ij}\d(t-t')\d({\bf x}-{\bf x'})\;,
\eeq 
and $\s$ is a non-abelian conductivity. The sphaleron rate can then be 
written as\cite{moo00},
\beq
\G \simeq (14.0 \pm 0.3) {1\over \s} (\a_w T)^5\;.
\eeq
A comparison with two lattice simulations is shown in fig.~\ref{boed_fig}.
From this one derives that $B$- and $L$-violating processes are in thermal
equilibrium for temperatures in the range
\begin{equation}
T_{EW} \sim 100\ \mbox{GeV} < T < T_{SPH} \sim 10^{12}\ \mbox{GeV}\;.
\end{equation}

Sphaleron processes have a profound effect on the generation of the
cosmological baryon asymmetry, in particular in connection with the dominant
lepton number violating interactions between lepton and Higgs fields,
\begin{equation}\label{dl2}
\cl_{\Delta L=2} ={1\over 2} f_{ij}\ l^T_{Li}\f C l_{Lj}\f 
                  +\mbox{ h.c.}\;.\label{intl2}
\end{equation}
Such an interaction arises in particular from the exchange of heavy Majorana
neutrinos. In the Higgs phase of the standard
model, where the Higgs field acquires a vacuum expectation value, it gives
rise to Majorana masses of the light neutrinos $\n_e$, $\n_\m$ and $\n_\t$.   

One may be tempted to conclude from eq.~(\ref{sphal1}) that any
$B+L$ asymmetry generated before the electroweak phase transition,
i.e., at temperatures $T>T_{EW}$, will be washed out. However, since
only left-handed fields couple to sphalerons, a non-zero value of
$B+L$ can persist in the high-temperature, symmetric phase in case
of a non-vanishing $B-L$ asymmetry. An analysis of the chemical potentials
of all particle species in the high-temperature phase yields a
relation between the baryon asymmetry $Y_B = (n_B-n_{\bar{B}})/s$ and the 
corresponding $B-L$ and $L$ asymmetries 
$Y_{B-L}$ and $Y_L$, respectively\cite{ht90},
\beq\label{basic}
Y_B\ =\ a\ Y_{B-L}\ =\ {a\over a-1}\ Y_L\;.
\eeq
The number $a$ depends on the other processes which are in thermal 
equilibrium. If these are all standard model interactions one has 
$a=28/79$. If instead of the Yukawa interactions
of the right-handed electron the $\D L=2$ interactions (\ref{dl2}) are in
equilibrium one finds $a=-2/3$ ~\cite{bp00}.

From eq.~(\ref{basic}) one concludes that the cosmological baryon asymmetry,
if generated before the electroweak transition,
requires also a lepton asymmetry, and therefore lepton number violation.
This leads to an intriguing interplay
between Majorana neutrinos masses, which are generated by the lepton-Higgs
interactions (\ref{dl2}), and the baryon asymmetry:
lepton number violating interactions are needed in order to generate a
baryon asymmetry; however, they have to be sufficiently weak, so that they
fall out of thermal equilibrium at the right time and a generated asymmetry
can survive until today.

The connection between baryon and lepton number at high temperatures leads to 
very attractive new mechanisms to generate the cosmological baryon asymmetry.
This includes large lepton number violating classical fields generated during
an inflationary phase and the out-of-equilibrium decays of heavy Majorana
neutrinos. In the following we shall compare these two versions of leptogenesis
which generically make different predictions for neutrino masses.

\section{Affleck-Dine leptogenesis}

In supersymmetric theories the $\D L = 2$ lepton-Higgs interactions (\ref{dl2})
are contained 
in the superpotential
\beq\label{sdl2}
W = {1\over 2M_i} \f L_i \f L_i\;,
\eeq
where $H_2$ and $L_i$, $i=1\ldots 3$ denote Higgs and lepton superfields in a
particular basis, respectively. The supersymmetric standard model possesses
the D-flat direction\cite{my94}
\bea
\f = \left(\begin{array}{c} \vf \\ 0 \end{array}\right)\;, \quad
L_1 = \left(\begin{array}{c} 0 \\ \vf \end{array}\right)\;, 
\eea
whose flatness is lifted by the non-renormalizable interaction (\ref{sdl2}) and
by by various supergravity and finite-temperature corrections. 
Recently, leptogenesis based on this flat direction has been studied in
detail\cite{afx00,fhy01}. The complete scalar potential during an inflationary
phase with Hubble parameter $H_I$ is given by\cite{fhy01}
\bea
V(\vf,T) &=& \mu(T)^2 |\vf|^2 + a_g \a_s^2 T^4 \ln{{|\vf|^2\over T^2}}
              +{m_{3/2}\over 8M_1}\left(a_m \vf^4 + c.c.\right)
              +{1\over 4M^2}|\vf|^6 \nonumber\\
&& - c_I H_I^2 |\vf|^2 + {H_I\over 8M_1}\left(a_H\vf^4 + c.c.\right)\;.
\eea
Here $a_g = \co(1)$ is a parameter which depends on the particle content
of the theory, $m_{3/2}a_m$ reflects supersymmetry breaking at $T=0$,
whereas $c_I H_I^2$ and $H_I a_H$ are due to the breaking of supersymmetry 
during inflation. Note, that the coefficient $c_I$ has to be positive
for Affleck-Dine leptogenesis to work whereas the simplest canonical K\"ahler
potential would yield the opposite sign for $c_I$! 
\begin{figure}[t]
\begin{center}
\epsfig{file=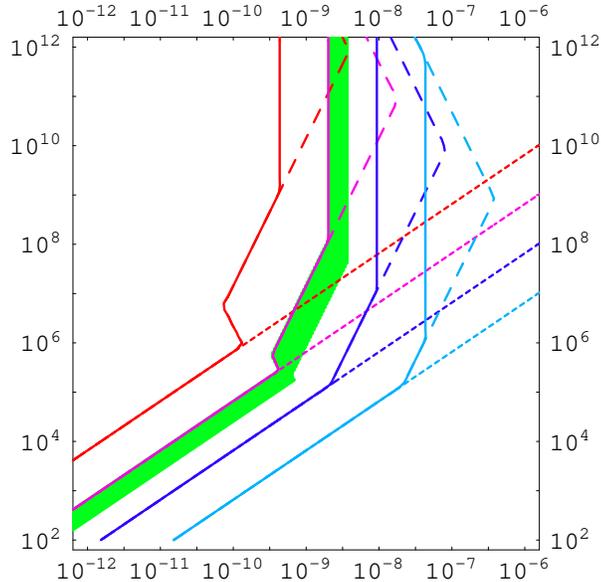,width=8cm,clip=}
\caption{Contour plot for the baryon asymmetry $Y_B=n_B/s$ in the 
$m_{\nu_1}\mbox{[eV]}(horizontal)-T_R\mbox{[GeV]}$(vertical)-plane. The full 
lines correspond to
$Y_B = 10^{-9}$, $10^{-10}$, $10^{-11}$ and $10^{-12}$ from left to right.
The short-dashed lines represent the results obtained negleting thermal 
effects; the long-dashed lines give the baryon asymmetries if only thermal 
masses are included. The shaded region corresponds to the observed baryon 
asymmetry\protect\cite{fhy01}.}
\label{fhy1_fig}
\end{center}
\end{figure}

The assumed phase difference between the complex numbers $a_m$ and $a_H$ is 
responsible for the generation of a lepton asymmetry. During the inflationary
phase the field $\vf$ is driven to one of the discrete minima
\beq
|\vf| = \vf_0 \sim (MH_I)^{1/2}\;,\quad
arg(\vf) = {1\over 4} ( - arg(a_H) + (2n+1)\pi)\;,\;n=0\ldots 3\;.
\eeq
After inflation, when $\m(T)^2$ starts to be the dominant mass term, the 
homogeneous field $\vf$
begins coherent oscillations which eventually lead to an asymmetry in the
number densities for ordinary leptons and scalar leptons and, finally,
via the sphaleron processes to an asymmetry in baryon number. It is remarkable
that the final baryon asymmetry is rather insensitive to the reheating
temperature after inflation in the range $T_R \simeq 10^5 \ldots 10^{12}$GeV.
A detailed calculation yields\cite{fhy01},
\beq
{n_B\over s} \simeq 10^{-11} \d_{eff} 
 \left({m_{\n_1}\over 10^{-8}\mbox{eV}}\right)^{-3/2}
\left({m_{3/2}|a_m|\over 1 \mbox{TeV}}\right)\;,
\eeq
i.e., the baryon asymmetry is determined by the lightest neutrino mass and
the strength of supersymmetry breaking. For an effective $C\!P$-violating angle
$\d_{eff} \simeq 0.1 \ldots 1$ and $m_{3/2}|a_m| \simeq 1$TeV, the observed
baryon asymmetry is obtained for an `ultralight' neutrino,
\beq
m_{\n_1} \simeq (0.1 - 1)\times 10^{-9} \mbox{eV}\;.
\eeq
This small neutrino mass reflects the required flatness of the D-flat direction
for successful Affleck-Dine leptogenesis.

\begin{figure}[t]
\begin{center}
\epsfig{file=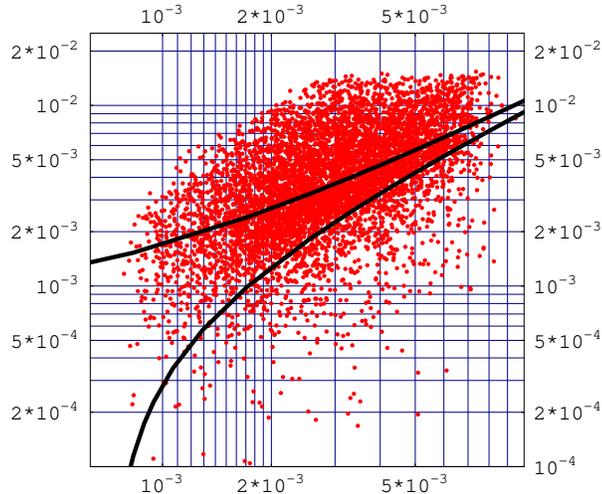,width=8cm,clip=}
\caption{$m_{\nu e \nu e}$[eV] as function of 
$\sin^2{\theta_{sol}}\sqrt{\Delta m^2_{sol}}$[eV] for the large angle MSW 
solution and $|U_{e3}|<0.10$\protect\cite{fhy01}.}
\label{fhy2_fig}
\end{center}
\end{figure}

The prediction of an ultralight neutrino has direct implications for 
neutrinoless double beta decay. The effective neutrino mass which determines
the decay rate is given by $m_{\nu_e \nu_e}=|\sum_i U_{ei}^2m_{\nu_i}|$, where
$m_{\nu_i}$ are the neutrino masses and $U_{\alpha i}$ the neutrino mixing
matrix, respectively. For hierarchical neutrinos with an ultralight lowest
state $m_{\nu_e \nu_e}$ can be expressed in terms of the oscillation parameters
determined from the solar and the atmospheric neutrino deficits,
\bea
m_{\nu_e \nu_e}= \left|\left(1-|U_{e3}|^2\right)
\sin^2{\theta_{sol}}\sqrt{\Delta m^2_{sol}} + 
|U_{e3}|^2 e^{i\alpha}\sqrt{\Delta m^2_{atm}}\right|\;.
\eea
It is remarkable that for the large angle MSW solution $m_{\nu_e \nu_e}$
reaches the value of $m_{\nu_2} \simeq \sqrt{\Delta m^2_{sol}} \simeq
3\times 10^{-3} - 10^{-2}$[eV] (cf.~fig.~\ref{fhy2_fig}), which may be 
accessible in future neutrinoless double beta decay experiments.

\section{Thermal Leptogenesis}

Here one starts from a thermal distribution of heavy Majorana fermions
which have $C\!P$ violating decay modes into standard model leptons. The
natural candidates are the three right-handed neutrinos $\n_{Ri}$, 
$i=1\ldots 3$, whose interactions are described by the lagrangian,
\beq
\cl = \Bar{l}_L \f^*\l^* \n_R - {1\over 2}\Bar{\n}_R^cM\n_R + h.c.
\eeq
$M$ is a Majorana mass matrix. The vacuum expectation value of the Higgs 
field $\VEV{\f}=v$ generates the Dirac mass term $m_D=h_{\n}v$
which is assumed to be small compared to the Majorana mass $M$.  
This yields light and heavy neutrino mass eigenstates according to the seesaw 
mechanism\cite{seesaw},
  \beq
     \n\simeq V_{\nu}^T\n_L+\n_L^c V_{\nu}^*\quad,\qquad
     N\simeq\n_R+\n_R^c\, ,
  \eeq
with masses
  \beq
     m_{\n}\simeq- V_{\nu}^Tm_D^T{1\over M}m_D V_{\nu}\,
     \quad,\quad  m_N\simeq M\; ,
     \label{seesaw}
  \eeq
where $V_{\nu}$ is the neutrino mixing matrix.

We shall restrict our discussion to the case of hierarchical Majorana neutrino
masses, $M_1 \ll M_2,M_3$. The baryon asymmetry is then determined by
the $C\!P$ violating decays of the lightest Majorana neutrino
$N_1=\n_{R1}+\n_{R1}^c\equiv N$,
\beq\label{gcp} 
\G(N\rightarrow l\phi) = {1\over 2}(1+\ve)\G\;,\quad
\G(N\rightarrow \bar{l}\bar{\phi})={1\over 2}(1-\ve)\G\;.
\eeq
Here $\G=(\l^\dagger\l)_{11}M/(8\pi)$ is the total decay width, and the
parameter $\ve \ll 1$ measures the 
amount of $C\!P$ violation. The generation of the baryon asymmetry takes
place at a temperature $T\sim  M = M_1 \ll M_2,M_3$. It is therefore
convenient to describe the system by an effective lagrangian where the two
heavier neutrinos have been integrated out,
\bea\label{lint}
\cl &=& \Bar{l}_{Li} \f^*\l^*_{i1} N + 
      N^T \l_{i1} C l_{Li}\f - {1\over 2}M N^T C N \NO\\
&& + {1\over 2}\h_{ij} l_{Li}^T\f C l_{Lj}\f
   + {1\over 2}\h^*_{ij} \Bar{l}_{Li}\f^* C \Bar{l}_{Lj}^T\f^*\;,
\eea
with
\beq
\h_{ij}=\sum_{k=2}^3\l_{ik}{1\over M_k}\l^T_{kj}\;.
\eeq
The $C\!P$ asymmetry $\ve$ arises from one-loop vertex and self-energy 
corrections\cite{fps95,crv96,bp98}. It can be expressed in a
compact form as\cite{bf00}
\beq\label{cpa}
\ve = {3\over 16\p}{\mbox{Im}(\l^\dg\h\l^*)_{11}\over (\l^\dg\l)_{11}}M\; .
\eeq

Given the Dirac and Majorana neutrino mass matrices the $C\!P$ asymmetry
$\ve$ and, consequently, the baryon asymmetry are determined. Consider
as an example a pattern of fermion masses based on the group
$SU(5)_{GUT}\times U(1)_F$, where $U(1)_F$ is a spontaneously broken 
generation symmetry. The Yukawa couplings 
arise from non-renormalizable interactions with a gauge singlet field $\F$ 
which acquires a vacuum expectation value\cite{fn79},
\beq\label{lfn}
\l_{ij} = g_{ij} \left({\VEV\F\over \L}\right)^{Q_i + Q_j}\;.
\eeq
Here $g_{ij}$ are couplings $\co(1)$ and $Q_i$ are the $U(1)_F$ charges of the
various fermions, with $Q_{\F}=-1$. The interaction scale $\L$ is
usually chosen to be very large, $\L > \L_{GUT}$. An example of possible 
charges $Q_i$ is given in table~1.
\begin{table}[b]
\caption{Chiral charges of charged and neutral leptons with
   $SU(5)_{GUT}\times U(1)_F$ symmetry.}
\begin{center}
\begin{tabular}{c|ccccccccc}\hline \hline
$\j_i$       & $ e^c_{R3}$ & $ e^c_{R2}$  & $ e^c_{R1}$  & $ l_{L3}$    & 
$ l_{L2}$    & $ l_{L1}$   & $ \n^c_{R3}$ & $ \n^c_{R2}$ & $ \n^c_{R1}$ 
\\\hline
$Q_i$  & 0 & 1 & 2 & $0$ & $0$ & $1$ & 0 & $1$ & $2$ \\ \hline\hline
\end{tabular}
\end{center}
\end{table}
The assignment of the same charge to the lepton doublets of the second and 
third generation leads to a neutrino mass matrix of the form\cite{ys99,ram99}, 
\beq\label{matrix}
m_{\n_{ij}} \sim \left(\begin{array}{ccc}
    \e^2  & \e  & \e \\
    \e  & \; 1 \; & 1 \\
    \e  &  1  & 1 
    \end{array}\right) {v_2^2\over \VEV R}\;.
\eeq
This structure immediately yields a large $\n_\m -\n_\t$ mixing angle. The
phenomenology of neutrino oscillations depends on the unspecified coefficients
which are $\co(1)$. The flavour mixing parameter $\e$ is chosen to be
$\VEV\F/\L = \e  \sim 1/ 17$, 
which corresponds to the mass ratio $m_\m/m_\t$.

One easily verifies that the mass ratios of heavy and light Majorana 
neutrinos are given by 
\bea
M_1 : M_2  : M_3  \sim \e^4 : \e^2 : 1\;,\quad
m_1 : m_2  : m_3  \sim \e^2 : 1 : 1\;.
\eea
The masses of the two eigenstates $\n_2$ and $\n_3$ depend on the unspecified 
factors ${\co(1)}$. They are therefore consistent with the mass differences 
$\D m^2_{\n_1 \n_2}\simeq 10^{-5} - 10^{-3}$~eV$^2$ inferred from the 
large angle MSW solution of the solar neutrino problem  and 
$\D m^2_{\n_2 \n_3}\simeq 10^{-3} - 10^{-2}$~eV$^2$ associated 
with the atmospheric neutrino deficit\cite{rpp00}. In the following we
we shall use for numerical estimates the geometric mean 
of the neutrino masses of the second and third family,
$\Bar{m}_\n=(m_{\n_2}m_{\n_3})^{1/2} \sim 10^{-2}$~eV. 
The choice of the charges in table~1 corresponds to large Yukawa couplings
of the third generation. For the mass of the heaviest Majorana neutrino
one then finds
\beq
M_3\ \sim\ {v^2\over\Bar{m}_\n}\ \sim\ 10^{15}\ \mbox{GeV}\;,
\eeq
which implies that $B-L$ is broken at the unification scale $\L_{GUT}$.

The $C\!P$ asymmetry $\ve$ (\ref{cpa}) leads to a lepton asymmetry in the
course of the cosmological evolution\cite{kt90},
  \beq\label{basym}
    Y_L\ =\ {n_L-n_{\Bar{L}}\over s}\ =\ \k\ {\ve\over g_*}\;.
  \eeq
  Here the factor $\k<1$ represents the effect of washout
  processes. In order to determine $\k$ one has to solve the full
  Boltzmann equations\cite{lut92,plu97}. 
  Important processes are the $\D L=2$ lepton Higgs scatterings
  mediated by heavy neutrinos since
  cancellations between on-shell contributions to these scatterings
  and contributions from neutrino decays and inverse decays ensure
  that no asymmetry is generated in thermal equilibrium.
  Further, due to the large top-quark Yukawa coupling one has to take
  into account neutrino top-quark scatterings mediated by 
  Higgs bosons\cite{lut92,plu97}. 
\begin{figure}[t]
    \mbox{ }\hfill
    \epsfig{file=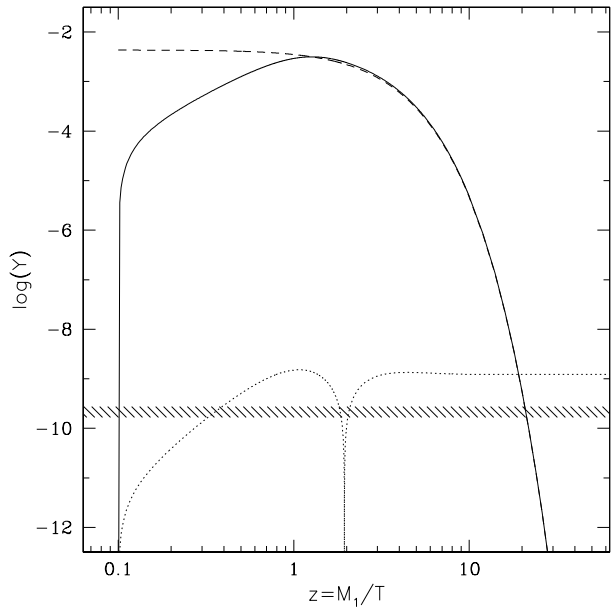,width=8.2cm}
    \hfill\mbox{ }
    \caption{Time evolution of the heavy neutrino number density and the
     lepton asymmetry for the $SU(5)_{GUT}\times U(1)_F$ model. 
     The solid line shows the solution of the Boltzmann equation for the 
     right-handed neutrinos, while the corresponding equilibrium 
     distribution is represented by the dashed line.
     The absolute value of the lepton asymmetry $Y_L$ 
     is given by the dotted line and the hatched area shows the
     lepton asymmetry corresponding to the observed baryon asymmetry
     \protect\cite{bp00}.\label{asyB_fig}}
\end{figure} 
These processes are of crucial importance for leptogenesis, since
they can create a thermal population of heavy neutrinos at high
temperatures $T>M$.

In order to obtain the baryon asymmetry predicted by the model of neutrino
masses described above we first have to evaluate the $C\!P$ asymmmetry.
Since for the Yukawa couplings only the powers in $\e$ are known, we will
also obtain the $C\!P$ asymmetry and the corresponding baryon asymmetry
to leading order in $\e$, i.e., up to unknown factors ${\cal O}(1)$. 
Note, that for models with a $U(1)_F$ generation symmetry the baryon
asymmetry is `quantized', i.e., changing the $U(1)_F$ charges will change
the baryon asymmetry by powers of $\e$\cite{by99}.
One easily obtains from eqs.~(\ref{cpa}), (\ref{lfn}) and the table\cite{by99},
\beq
  \ve\ \sim\ {3\over 16\pi}\ \e^4\;.
  \label{su5epsilon}
\eeq 
Using $\e^2 \sim 1/300$  and $g_* \sim 100$ 
one then obtains the baryon asymmetry,
\beq\label{est1}
Y_B \sim \k\ 10^{-8}\;.
\eeq
For $\k \sim 0.1\ldots 0.01$ this is indeed the correct order of magnitude!
The baryogenesis temperature is given by the mass of the lightest of the
heavy Majorana neutrinos,
\beq
T_B \sim M_1 \sim \e^4 M_3 \sim 10^{10}\ \mbox{GeV}\;.
\eeq
For this model, where the $C\!P$ asymmetry is determined by the mass 
hierarchy of light and heavy Majorana neutrinos, baryogenesis has been 
studied in detail in ref\cite{bp96}. It is remarkable that the observed
baryon asymmety is obtained without any fine tuning of parameters, if
$B-L$ is broken at the unification scale $\L_{GUT}$.
The generated baryon asymmetry does not depend on the flavour mixing of the 
light neutrinos, in particular the $\n_\m-\n_\t$ mixing angle. 

The solution of the full Boltzmann equations is shown in fig.~\ref{asyB_fig}. 
The initial condition at a temperature $T \sim 10 M_1$ is 
chosen to be a state without heavy neutrinos. The Yukawa interactions are 
sufficient to bring the heavy neutrinos into thermal equilibrium. 
At temperatures $T\sim M_1$ the familiar out-of-equilibrium decays set in,
which leads to a non-vanishing baryon asymmetry. The final asymmetry agrees 
with the estimate (\ref{est1}) for $\k \sim 0.1$. The dip in 
fig.~\ref{asyB_fig} is due to a change of sign in the lepton asymmetry at 
$T \sim M_1$. 

The final baryon asymmetry is usually obtained from the final
lepton asymmetry by means of eq.~(\ref{basic}). Note, that this is conceptually
not correct! The sphaleron processes are in equilibrium whereas the heavy
neutrino decays out of equilibrium. Hence, any generated asymmetry is
immediately distributed among other degrees of freedom in the plasma.
This effect reduces the generated baryon asymmetry by a factor 
${\co(1)}$\cite{bp01}.

\section{Towards the Theory of Leptogenesis}

The generation of a baryon asymmetry is an out-of-equilibrium process 
which is generally treated by means of Boltzmann equations. 
A shortcoming of this
approach is that the Boltzmann equations are classical equations for the time 
evolution of phase space distribution functions. On the contrary, the
involved collision terms are $S$-matrix elements which involve
quantum interferences of different amplitudes in a crucial manner. Clearly,
a full quantum mechanical treatment is highly desirable. This is also
required in order to justify the use of the Boltzmann equations and to 
determine the size of corrections.

All information about the time evolution of a system is contained in the
time dependence of its Green functions\cite{kb62,ke64}
which can be determined by means of Dyson-Schwinger equations. Originally
these techniques were developed for non-relativistic many-body problems. More 
recently, they have also been applied to transport phenomena in 
nuclear matter\cite{mh94}, 
the electroweak plasma\cite{ri98,jkp00} and the QCD plasma\cite{bi99}. 
Alternatively, one may study the time evolution of density 
matrices\cite{jmy98,abv98}. In the following we shall describe how the
Green function technique can be used to obtain a systematic perturbative 
expansion around a solution of the Boltzmann equations\cite{bf00}.
  
The time evolution of an arbitrary multi-particle lepton-Higgs system can be 
studied by means of the Green functions of lepton and Higgs fields. For the 
heavy Majorana neutrino one has
\beq
iG_{\a\b}(x_1,x_2)=\mbox{Tr}\left(\r \hat{T} N_\a(x_1)N_\b(x_2)\right)\;,
\eeq
where $\hat{T}$ denotes the time ordering, $\r$ is the density matrix of 
the system,
the trace extends over all states, and the time coordinates $t_1$ and $t_2$
lie on an appropriately chosen contour $C$ in the complex plane\cite{lb96}. 
$G(x_1,x_2)$ can be written as a sum of two parts,
\beq\label{gn}
G(x_1,x_2)=\Q(t_1-t_2) G^>(x_1,x_2)+\Q(t_2-t_1) G^<(x_1,x_2)\; .
\eeq  

For a system in thermal equilibrium at a temperature $T=1/\b$ the density 
matrix is $\r=\exp{(-\b H)}$, where $H$ is the Hamilton operator. In this
case the Green function only depends on the difference of coordinates and it 
is convenient to work with the Fourier transform $G(p)$.
The contour $C$ can be chosen as a sum of two branches, $C=C_1\cup C_2$, which
lie above and below the real axis. The time coordinates are real 
and associated with one of the two branches. Correspondingly, the Green 
function becomes a $2\times 2$ matrix,
\bea\label{gmatrix}
G(p)=\left(
\begin{array}{cc}
G^{11}(p) & G^{12}(p)  \\[1ex]
G^{21}(p) & G^{22}(p) 
\end{array}
\right)\; . 
\eea
The off-diagonal terms are given by
\beq
G^{12}(p)=G^<(p)\;, \quad G^{21}(p)=G^>(p)\;,
\eeq
the diagonal terms of the matrix (\ref{gmatrix}) are the familiar causal and 
anti-causal Green functions. 
The free Green functions are explicitly given by
\bea
iG^>(p)&=&\left(\Q(p_0)
                -\Q(p_0)f_N(E)-\Q(-p_0)f_{\bar{N}}(E)\right)\r_N(p)\;,\\ 
iG^<(p)&=&\left(\Q(-p_0)  
                 -\Q(p_0)f_N(E)-\Q(-p_0)f_{\bar{N}}(E)\right)\r_N(p)\;,
\eea
with the spectral density
\beq\label{nspectral}
\r_N(p) = 2\p (\slash p + M)C^{-1}\d(p^2-M^2)\;,
\eeq 
and the Fermi-Dirac distribution functions
\beq\label{fn}
f_N(E)=f_{\bar{N}}(E)={1\over e^{\b E} + 1}\ , \quad E=\sqrt{M^2+p^2}\; .
\eeq
Since $N(x)$ is a Majorana field one has $f_N=f_{\bar{N}}$, and the charge 
conjugation matrix $C$ occurs in the spectral density (\ref{nspectral}).
The Green functions $S(x_1,x_2)$ and $\D(x_1,x_2)$ for the lepton doublets
and the Higgs doublet also depend on chemical potentials. The 
Schwinger-Dyson equations for the Green functions $G^>(x_1,x_2)$, 
$G^<(x_1,x_2)$, etc. are usually referred to as Kadanoff-Baym equations.

\begin{figure}
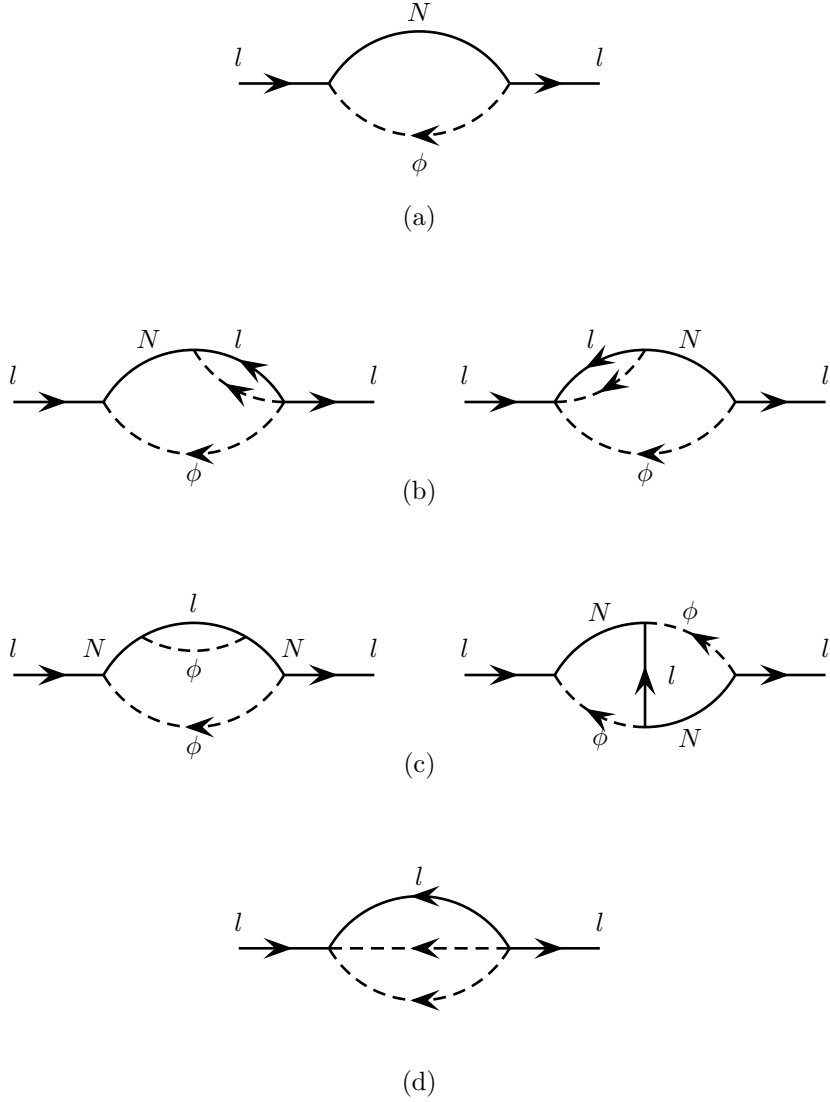

\input{fig2a.tex}
\input{fig2b.tex}
\input{fig2c.tex}
\input{fig2d.tex}
\caption{One- and two-loop self energies for the lepton doublet. 
\label{seld}}
\end{figure}

For processes where the overall time evolution is slow compared to relative
motions the Kadanoff-Baym equations can be solved in a derivative expansion.
One considers the Wigner transform for $G(x_1,x_2)$,
\beq
G(x,p)=\int d^4y\ e^{ipy}\ G\left(x+{y\over 2},x-{y\over 2}\right)\;,
\eeq
and for $S(x,p)$ and $\D(x,p)$, respectively. To leading order in
the derivative expansion the Kadanoff-Baym equations
become local in the space-time coordinate
$x$. Keeping to zeroth order only the on-shell part of retarded and advanced
Green functions and self-energies one obtains the equations
\bea
C({i\over 2}\slpa + \slash p - M)G^>(x,p)&=&
C({i\over 2}\slpa + \slash p - M)G^<(x,p)\NO\\
&=&{1\over 2}\left(\S^>(x,p)G^<(x,p)-\S^<(x,p)G^>(x,p)\right)\;,\label{kbN}\\
({i\over 2}\slpa + \slash k )S^>(x,p)&=&
({i\over 2}\slpa + \slash k )S^<(x,p)\NO\\
&=&{1\over 2}\left(\P^>(x,k)S^<(x,k)-\P^<(x,p)S^>(x,p)\right)\;.\label{kbl}
\eea
The corresponding self-energies are shown in fig.~(\ref{seld}).
Solutions of these equations yield the first terms for the non-equilibrium 
Green functions $G^>(x,p)$...$S^<(x,k)$ in an expansion involving off-shell
effects and space-time variations which include `memory effects'. 

Given the lepton self-energies we can now look for solutions of the 
equations (\ref{kbN}) and (\ref{kbl}). A straightforward
calculation shows that the right-hand side of these equations vanishes
for equilibrium Green functions and self-energies. Since leptogenesis is
a process close to thermal equilibrium the solutions should be linear in 
the deviations,
\bea
\d G(t,p) &=& G^>(t,p)-G^>_{eq}(p) = G^<(t,p)-G^<_{eq}(p)\;,\\
\d S(t,p) &=& S^>(t,p)-S^>_{eq}(p) = S^<(t,p)-S^<_{eq}(p)\;.
\eea
One may also expect that $\d G(t,p)$ and $\d S(t,p)$ can be obtained from the
equilibrium Green functions by a small change of the distribution functions,
\bea
i\d G(x,p) = -\d f_N(x,p)\r_N(p)\;, \quad
i\d S(x,k) = -\e(k_0) \d f_l(x,k)\r_l(k)\;.
\eea
This ansatz indeed reduces the matrix equations (\ref{kbN}) and (\ref{kbl})
to the system of ordinary differential equations  
\bea
E\der t \d f_N(t,p) &=& 
-E\der t f_N(p) - 2 (\l^\dg\l)_{11} \int d\F_{\bar{1}\bar{2}}(p)
\d f_N(t,p) p\cdot p_1\;,\label{diffn}\\
g_l k \der t \d f_l(t,k) &=&  
{3\over 8\p}\mbox{Im}(\l^\dg\h\l^*)_{11}M \int d\F_{\bar{1}2}(k)
\d f_N(t,p_1) k\cdot p_1\ \NO\\
&&-2 (\l^\dg\l)_{11}\int d\F_{1\bar{2}}(k) 
\left(\d f_l(t,k)f_\f(p_1)+f_l(k)\d f_\f(t,p_1)\right) k\cdot p_2 \NO\\
&&-6 {(\l^\dg\l)^2_{11}\over M^2} \int d\F_{1\bar{2}\bar{3}}(k)\Big(
2(\d f_l(t,k)f_\f(p_1)+f_l(k)\d f_\f(t,p_1)\NO\\
&&\hspace{2cm}+\d f_l(t,p_2)f_\f(p_3)+f_l(p_2)\d f_\f(t,p_3))
k\cdot p_2 \NO\\
&&\hspace{2cm}+(\d f_l(t,k)f_l(p_1)+f_l(k)\d f_l(t,p_1)\NO\\
&&\hspace{2.5cm}+2\d f_\f(t,p_2)f_\f(p_3)) k\cdot p_1\Big)\ .\label{diffl}
\eea
Here $g_l=6$ is the number of `internal' degrees of freedom for three 
generations of lepton doublets, $d\F_{\bar{1}\bar{2}}(p)$ etc. denote
phase space integrations, and $f_N(p)$, $f_l(p)$ and $f_\f(p)$ are
the equilibrium distributions of heavy neutrinos, lepton doublets and
Higgs doublet, respectively.

Eqs.~(\ref{diffn}) and (\ref{diffl}) are the Boltzmann equations for 
the distributions functions
of heavy neutrinos and lepton doublets with the correct matrix elements for
decays, $2\rightarrow 2$ processes and $C\!P$ violation. The
effect of the Hubble expansion is included by means of the substitution
$\pa/\pa t \rightarrow \pa/\pa t - H p \pa/\pa p$. Integration over
momenta then yields the more familiar form of the Boltzmann equations for the 
number densities. Note, that also the equilibrium distributions are time 
dependent since the temperature varies with time. 
The first term in eq.~(\ref{diffl}) drives the generation of a 
lepton asymmetry; the remaining terms tend to wash out an existing asymmetry. 
 
We conclude that a solution of the Boltzmann equations generates a
solution of the Kadanoff-Baym equations to leading order in the
expansion described above. Various corrections such as relativistiv effects,
`memory effects', off-shell effects and deviations from kinetic equilibrium
can now be systematically studied. By means of such an analysis one can
obtain quantitative constraints on the parameters $M$, $(\l^\dg\l)_{11}$ 
and $\ve$, which relate the cosmological baryon asymmetry and neutrino 
properties.

\section{Outlook}

Detailed studies of the thermodynamics of the electroweak interactions at
high temperatures have shown that in the standard model and most of its
extensions the electroweak transition is too weak to affect the 
cosmological baryon asymmetry. Hence, one has to search for baryogenesis
mechanisms above the Fermi scale. 

Due to sphaleron processes baryon number and lepton number are related
in the high-temperature symmetric phase of the standard model. As a
consequence, the cosmological baryon asymmetry is related to neutrino
properties. Generically, baryogenesis requires lepton number violation, which 
occurs in extensions of the standard model with right-handed neutrinos and
Majorana neutrino masses. In detail the relations between $B$, $L$ and $B-L$
depend on all other processes taking place in the plasma, and therefore
also on the temperature.  

Although lepton number violation is needed in order to obtain a baryon
asymmetry, it must not be too strong since otherwise any baryon and lepton
asymmetry would be washed out. Hence, leptogenesis leads to upper 
and lower bounds on the masses of the light and heavy Majorana neutrinos,
respectively.

Different realizations of leptogenesis imply different constraints on
neutrino masses. Affleck-Dine leptogenesis based on the lepton-Higgs
D-flat direction predicts an ultralight neutrino with mass 
$m_{\n_1} \sim 10^{-9}$~eV. Also thermal leptogenesis requires neutrino
masses significantly below 1~eV. However, in this case the neutrino masses
may be as large as $\sim 10^{-2}$~eV, which corresponds to the mass 
splitting indicated by the atmospheric neutrino anomaly.   
It is very remarkable that the observed baryon asymmetry 
$n_B/s \sim 10^{-10}$ is naturally explained by the decay of heavy Majorana 
neutrinos, with $B-L$ broken at the unification
scale $\Lambda_{\mbox{\scriptsize GUT}}\sim 10^{16}\;$GeV, and in accord
with present experimental indications for neutrino masses.

Further work is needed to develop a full quantum mechanical description of 
leptogenesis which goes beyond the Boltzmann equations. Also important is
the connection with other lepton and quark flavour changing processes in
the context of unified theories. Finally, the realization of the rather large
baryogenesis temperature $T_B \sim 10^{10}$~GeV in models of inflation
should have implications for dark matter and the cosmic microwave background.

\end{document}